\documentclass[
reprint,
amsmath,amssymb,
aps,
]{revtex4-2}

\PassOptionsToPackage{hyphens}{url}\usepackage{graphicx}
\usepackage{dcolumn}
\usepackage{bm}
\usepackage{braket}
\usepackage{changepage}
\usepackage{tikz}
\usetikzlibrary{decorations.pathreplacing,decorations.pathmorphing}
\usepackage{float}
\usepackage{url}
\usepackage{xurl}

\begin{document}
	
	\preprint{APS/1}
	
	\title{Using Universal Frame Randomization and Randomized Compilation to Mitigate Errors in Quantum Optimization}
	
	\author{Rachel E. Johnson}
	%\affiliation{University of Notre Dame}
	\affiliation{Lockheed Martin}
	\email{rachel.e1.johnson@lmco.com}

	\author{Joshua A. Job}
	\affiliation{Lockheed Martin}

	\author{Steve Adachi}
	\affiliation{Lockheed Martin}
	
	\date{\today}
	
	\begin{abstract}
		Error mitigation is essential for near-term quantum devices, and two promising techniques are universal frame randomization and Randomized Compilation. These methods insert random twirling gates into a circuit to reduce errors while preserving unitarity and depth. We apply universal frame randomization and Randomized Compilation to the quantum approximate optimization algorithm (QAOA) with $p=1$ on a superconducting quantum circuit system, demonstrating its potential to improve energy calculations. Specifically, we investigate the use of QAOA to calculate the lowest energy state of a frustrated Ising ring system and compare the results of randomized circuits generated using both techniques.  Our results show that both methods can mitigate errors, with expected extremal energy values of $5.25\pm0.145$ and $4.08\pm0.36$, for Randomized Compilation and universal frame randomization respectively, compared to $2.63\pm0.068$ without randomization and $5.676\pm0.006$ with a noiseless simulator.
	\end{abstract}
	
	\maketitle

	\section{Introduction}
	
	Quantum computing has the potential to revolutionize various fields, but it is hindered by the presence of noise in quantum systems. To mitigate this noise, techniques such as quantum error correction, suppression, and mitigation are employed \cite{qctrl-website} \cite{nielsen} \cite{rieffel}. Quantum error correction is impractical for current noisy intermediate-scale quantum (NISQ) devices due to high error rates \cite{Preskill_2018}. Therefore, in this paper we focus on error mitigation strategies, universal frame randomization and Randomized Compilation, which have been shown to reduce errors in quantum computations \cite{cai2023quantum}\cite{Wallman_2016}. In this paper, we explore the application of universal frame randomization and Randomized Compilation to the Quantum Approximate Optimization Algorithm (QAOA) for solving a frustrated Ising ring problem.
	
	\subsection{Universal Frame Randomization \& Randomized Compilation}
	Universal frame randomization and Randomized Compilation are similar error mitigation techniques that mitigate errors associated with operating two-qubit gates \cite {Wallman_2016}.
	
\begin{figure}[H]
		\centering
		\begin{tikzpicture}[scale=1.000000,x=1pt,y=1pt]
			\filldraw[color=white] (12.500000, 0.000000) rectangle (-62.500000, -120.000000);
			% Drawing wires
			% Line 2: a W 0 width=25
			\draw[color=black] (-50.000000,0.000000) -- (-50.000000,-108.000000);
			\draw[color=black] (-49.500000,-108.000000) -- (-49.500000,-120.000000);
			\draw[color=black] (-50.500000,-108.000000) -- (-50.500000,-120.000000);
			\draw[color=black] (-50.000000,0.000000) node[above] {$0$};
			% Line 3: b W 1 width=25
			\draw[color=black] (-25.000000,0.000000) -- (-25.000000,-108.000000);
			\draw[color=black] (-24.500000,-108.000000) -- (-24.500000,-120.000000);
			\draw[color=black] (-25.500000,-108.000000) -- (-25.500000,-120.000000);
			\draw[color=black] (-25.000000,0.000000) node[above] {$1$};
			% Line 4: c W 2 width=25
			\draw[color=black] (-0.000000,0.000000) -- (-0.000000,-108.000000);
			\draw[color=black] (0.500000,-108.000000) -- (0.500000,-120.000000);
			\draw[color=black] (-0.500000,-108.000000) -- (-0.500000,-120.000000);
			\draw[color=black] (-0.000000,0.000000) node[above] {$2$};
			% Done with wires; drawing gates
			% Line 6: a H
			\begin{scope}
				\draw[fill=white] (-50.000000, -12.000000) +(-45.000000:8.485281pt and 8.485281pt) -- +(45.000000:8.485281pt and 8.485281pt) -- +(135.000000:8.485281pt and 8.485281pt) -- +(225.000000:8.485281pt and 8.485281pt) -- cycle;
				\clip (-50.000000, -12.000000) +(-45.000000:8.485281pt and 8.485281pt) -- +(45.000000:8.485281pt and 8.485281pt) -- +(135.000000:8.485281pt and 8.485281pt) -- +(225.000000:8.485281pt and 8.485281pt) -- cycle;
				\draw (-50.000000, -12.000000) node {$H$};
			\end{scope}
			% Line 7: b H
			\begin{scope}
				\draw[fill=white] (-25.000000, -12.000000) +(-45.000000:8.485281pt and 8.485281pt) -- +(45.000000:8.485281pt and 8.485281pt) -- +(135.000000:8.485281pt and 8.485281pt) -- +(225.000000:8.485281pt and 8.485281pt) -- cycle;
				\clip (-25.000000, -12.000000) +(-45.000000:8.485281pt and 8.485281pt) -- +(45.000000:8.485281pt and 8.485281pt) -- +(135.000000:8.485281pt and 8.485281pt) -- +(225.000000:8.485281pt and 8.485281pt) -- cycle;
				\draw (-25.000000, -12.000000) node {$H$};
			\end{scope}
			% Line 8: c H
			\begin{scope}
				\draw[fill=white] (0.000000, -12.000000) +(-45.000000:8.485281pt and 8.485281pt) -- +(45.000000:8.485281pt and 8.485281pt) -- +(135.000000:8.485281pt and 8.485281pt) -- +(225.000000:8.485281pt and 8.485281pt) -- cycle;
				\clip (0.000000, -12.000000) +(-45.000000:8.485281pt and 8.485281pt) -- +(45.000000:8.485281pt and 8.485281pt) -- +(135.000000:8.485281pt and 8.485281pt) -- +(225.000000:8.485281pt and 8.485281pt) -- cycle;
				\draw (0.000000, -12.000000) node {$H$};
			\end{scope}
			% Line 10: a C b
			\draw (-50.000000,-36.000000) -- (-25.000000,-36.000000);
			\begin{scope}
				\draw[fill=white] (-50.000000, -36.000000) circle(3.000000pt);
				\clip (-50.000000, -36.000000) circle(3.000000pt);
				\draw (-53.000000, -36.000000) -- (-47.000000, -36.000000);
				\draw (-50.000000, -39.000000) -- (-50.000000, -33.000000);
			\end{scope}
			\filldraw (-25.000000, -36.000000) circle(1.500000pt);
			% Line 11: c X
			\begin{scope}
				\draw[fill=white] (0.000000, -36.000000) +(-45.000000:8.485281pt and 8.485281pt) -- +(45.000000:8.485281pt and 8.485281pt) -- +(135.000000:8.485281pt and 8.485281pt) -- +(225.000000:8.485281pt and 8.485281pt) -- cycle;
				\clip (0.000000, -36.000000) +(-45.000000:8.485281pt and 8.485281pt) -- +(45.000000:8.485281pt and 8.485281pt) -- +(135.000000:8.485281pt and 8.485281pt) -- +(225.000000:8.485281pt and 8.485281pt) -- cycle;
				\draw (0.000000, -36.000000) node {$X$};
			\end{scope}
			% Line 13: a G {T}
			\begin{scope}
				\draw[fill=white] (-50.000000, -60.000000) +(-45.000000:8.485281pt and 8.485281pt) -- +(45.000000:8.485281pt and 8.485281pt) -- +(135.000000:8.485281pt and 8.485281pt) -- +(225.000000:8.485281pt and 8.485281pt) -- cycle;
				\clip (-50.000000, -60.000000) +(-45.000000:8.485281pt and 8.485281pt) -- +(45.000000:8.485281pt and 8.485281pt) -- +(135.000000:8.485281pt and 8.485281pt) -- +(225.000000:8.485281pt and 8.485281pt) -- cycle;
				\draw (-50.000000, -60.000000) node {{T}};
			\end{scope}
			% Line 14: b Z c
			\draw (-25.000000,-60.000000) -- (-0.000000,-60.000000);
			\begin{scope}
				\draw[fill=white] (-25.000000, -60.000000) +(-45.000000:8.485281pt and 8.485281pt) -- +(45.000000:8.485281pt and 8.485281pt) -- +(135.000000:8.485281pt and 8.485281pt) -- +(225.000000:8.485281pt and 8.485281pt) -- cycle;
				\clip (-25.000000, -60.000000) +(-45.000000:8.485281pt and 8.485281pt) -- +(45.000000:8.485281pt and 8.485281pt) -- +(135.000000:8.485281pt and 8.485281pt) -- +(225.000000:8.485281pt and 8.485281pt) -- cycle;
				\draw (-25.000000, -60.000000) node {$Z$};
			\end{scope}
			\filldraw (-0.000000, -60.000000) circle(1.500000pt);
			% Line 18: a H
			\begin{scope}
				\draw[fill=white] (-50.000000, -84.000000) +(-45.000000:8.485281pt and 8.485281pt) -- +(45.000000:8.485281pt and 8.485281pt) -- +(135.000000:8.485281pt and 8.485281pt) -- +(225.000000:8.485281pt and 8.485281pt) -- cycle;
				\clip (-50.000000, -84.000000) +(-45.000000:8.485281pt and 8.485281pt) -- +(45.000000:8.485281pt and 8.485281pt) -- +(135.000000:8.485281pt and 8.485281pt) -- +(225.000000:8.485281pt and 8.485281pt) -- cycle;
				\draw (-50.000000, -84.000000) node {$H$};
			\end{scope}
			% Line 19: b H
			\begin{scope}
				\draw[fill=white] (-25.000000, -84.000000) +(-45.000000:8.485281pt and 8.485281pt) -- +(45.000000:8.485281pt and 8.485281pt) -- +(135.000000:8.485281pt and 8.485281pt) -- +(225.000000:8.485281pt and 8.485281pt) -- cycle;
				\clip (-25.000000, -84.000000) +(-45.000000:8.485281pt and 8.485281pt) -- +(45.000000:8.485281pt and 8.485281pt) -- +(135.000000:8.485281pt and 8.485281pt) -- +(225.000000:8.485281pt and 8.485281pt) -- cycle;
				\draw (-25.000000, -84.000000) node {$H$};
			\end{scope}
			% Line 20: c H
			\begin{scope}
				\draw[fill=white] (0.000000, -84.000000) +(-45.000000:8.485281pt and 8.485281pt) -- +(45.000000:8.485281pt and 8.485281pt) -- +(135.000000:8.485281pt and 8.485281pt) -- +(225.000000:8.485281pt and 8.485281pt) -- cycle;
				\clip (0.000000, -84.000000) +(-45.000000:8.485281pt and 8.485281pt) -- +(45.000000:8.485281pt and 8.485281pt) -- +(135.000000:8.485281pt and 8.485281pt) -- +(225.000000:8.485281pt and 8.485281pt) -- cycle;
				\draw (0.000000, -84.000000) node {$H$};
			\end{scope}
			% Line 22: a M
			\draw[fill=white] (-56.000000, -114.000000) rectangle (-44.000000, -102.000000);
			\draw[very thin] (-50.000000, -107.400000) arc (90:150:6.000000pt);
			\draw[very thin] (-50.000000, -107.400000) arc (90:30:6.000000pt);
			\draw[->,>=stealth] (-50.000000, -113.400000) -- +(80:10.392305pt);
			% Line 23: b M
			\draw[fill=white] (-31.000000, -114.000000) rectangle (-19.000000, -102.000000);
			\draw[very thin] (-25.000000, -107.400000) arc (90:150:6.000000pt);
			\draw[very thin] (-25.000000, -107.400000) arc (90:30:6.000000pt);
			\draw[->,>=stealth] (-25.000000, -113.400000) -- +(80:10.392305pt);
			% Line 24: c M
			\draw[fill=white] (-6.000000, -114.000000) rectangle (6.000000, -102.000000);
			\draw[very thin] (-0.000000, -107.400000) arc (90:150:6.000000pt);
			\draw[very thin] (-0.000000, -107.400000) arc (90:30:6.000000pt);
			\draw[->,>=stealth] (-0.000000, -113.400000) -- +(80:10.392305pt);
			% Done with gates; drawing ending labels
			% Done with ending labels; drawing cut lines and comments
			% Line 26: a b c @ 0 0 color=black style=dashed,rounded_corners=5pt
			\draw[draw opacity=1.000000,fill opacity=0.200000,color=black,dashed,rounded corners=5pt] (-62.500000,-3.000000) rectangle (12.500000,-21.000000);
			\draw[draw opacity=1.000000,fill opacity=0.200000,color=black,dashed,rounded corners=5pt] (-62.500000,-3.000000) rectangle (12.500000,-21.000000);
			% Line 27: a b c @ 1 1 color=black style=dashed,rounded_corners=5pt
			\draw[draw opacity=1.000000,fill opacity=0.200000,color=black,dashed,rounded corners=5pt] (-62.500000,-27.000000) rectangle (12.500000,-45.000000);
			\draw[draw opacity=1.000000,fill opacity=0.200000,color=black,dashed,rounded corners=5pt] (-62.500000,-27.000000) rectangle (12.500000,-45.000000);
			% Line 28: a b c @ 2 2 color=black style=dashed,rounded_corners=5pt
			\draw[draw opacity=1.000000,fill opacity=0.200000,color=black,dashed,rounded corners=5pt] (-62.500000,-51.000000) rectangle (12.500000,-69.000000);
			\draw[draw opacity=1.000000,fill opacity=0.200000,color=black,dashed,rounded corners=5pt] (-62.500000,-51.000000) rectangle (12.500000,-69.000000);
			% Line 29: a b c @ 3 3 color=black style=dashed,rounded_corners=5pt
			\draw[draw opacity=1.000000,fill opacity=0.200000,color=black,dashed,rounded corners=5pt] (-62.500000,-75.000000) rectangle (12.500000,-93.000000);
			\draw[draw opacity=1.000000,fill opacity=0.200000,color=black,dashed,rounded corners=5pt] (-62.500000,-75.000000) rectangle (12.500000,-93.000000);
			% Done with comments
		\end{tikzpicture}
		\caption{Circuit and cycles for randomized compilation on a simple quantum circuit. Each cycle is circled with a dashed line.}
		\label{fig:rc-example}
	\end{figure}
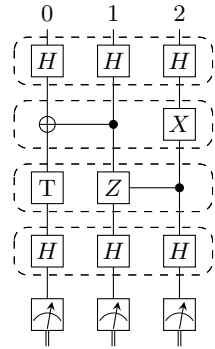

	To apply either of these techniques to a specific circuit, we start by drawing the circuit diagram and \textit{framing} different \textit{cycles} within the circuit. For example, in the circuit in Figure \ref{fig:rc-example}, we define 4 cycles and each cycle in this example contains no more than one gate operation for each qubit. 
	
\begin{figure}
		\centering
		\begin{tikzpicture}[scale=1.000000,x=1pt,y=1pt]
			\filldraw[color=white] (12.500000, 0.000000) rectangle (-62.500000, -312.000000);
			% Drawing wires
			% Line 2: a W 0 width=25
			\draw[color=black] (-50.000000,0.000000) -- (-50.000000,-300.000000);
			\draw[color=black] (-49.500000,-300.000000) -- (-49.500000,-312.000000);
			\draw[color=black] (-50.500000,-300.000000) -- (-50.500000,-312.000000);
			\draw[color=black] (-50.000000,0.000000) node[above] {$0$};
			% Line 3: b W 1 width=25
			\draw[color=black] (-25.000000,0.000000) -- (-25.000000,-300.000000);
			\draw[color=black] (-24.500000,-300.000000) -- (-24.500000,-312.000000);
			\draw[color=black] (-25.500000,-300.000000) -- (-25.500000,-312.000000);
			\draw[color=black] (-25.000000,0.000000) node[above] {$1$};
			% Line 4: c W 2 width=25
			\draw[color=black] (-0.000000,0.000000) -- (-0.000000,-300.000000);
			\draw[color=black] (0.500000,-300.000000) -- (0.500000,-312.000000);
			\draw[color=black] (-0.500000,-300.000000) -- (-0.500000,-312.000000);
			\draw[color=black] (-0.000000,0.000000) node[above] {$2$};
			% Done with wires; drawing gates
			% Line 7: a G {}
			\begin{scope}
				\draw[fill=white] (-50.000000, -12.000000) +(-45.000000:8.485281pt and 8.485281pt) -- +(45.000000:8.485281pt and 8.485281pt) -- +(135.000000:8.485281pt and 8.485281pt) -- +(225.000000:8.485281pt and 8.485281pt) -- cycle;
				\clip (-50.000000, -12.000000) +(-45.000000:8.485281pt and 8.485281pt) -- +(45.000000:8.485281pt and 8.485281pt) -- +(135.000000:8.485281pt and 8.485281pt) -- +(225.000000:8.485281pt and 8.485281pt) -- cycle;
				\draw (-50.000000, -12.000000) node {{}};
			\end{scope}
			% Line 8: b G {}
			\begin{scope}
				\draw[fill=white] (-25.000000, -12.000000) +(-45.000000:8.485281pt and 8.485281pt) -- +(45.000000:8.485281pt and 8.485281pt) -- +(135.000000:8.485281pt and 8.485281pt) -- +(225.000000:8.485281pt and 8.485281pt) -- cycle;
				\clip (-25.000000, -12.000000) +(-45.000000:8.485281pt and 8.485281pt) -- +(45.000000:8.485281pt and 8.485281pt) -- +(135.000000:8.485281pt and 8.485281pt) -- +(225.000000:8.485281pt and 8.485281pt) -- cycle;
				\draw (-25.000000, -12.000000) node {{}};
			\end{scope}
			% Line 9: c G {}
			\begin{scope}
				\draw[fill=white] (0.000000, -12.000000) +(-45.000000:8.485281pt and 8.485281pt) -- +(45.000000:8.485281pt and 8.485281pt) -- +(135.000000:8.485281pt and 8.485281pt) -- +(225.000000:8.485281pt and 8.485281pt) -- cycle;
				\clip (0.000000, -12.000000) +(-45.000000:8.485281pt and 8.485281pt) -- +(45.000000:8.485281pt and 8.485281pt) -- +(135.000000:8.485281pt and 8.485281pt) -- +(225.000000:8.485281pt and 8.485281pt) -- cycle;
				\draw (0.000000, -12.000000) node {{}};
			\end{scope}
			% Line 11: a H
			\begin{scope}
				\draw[fill=white] (-50.000000, -36.000000) +(-45.000000:8.485281pt and 8.485281pt) -- +(45.000000:8.485281pt and 8.485281pt) -- +(135.000000:8.485281pt and 8.485281pt) -- +(225.000000:8.485281pt and 8.485281pt) -- cycle;
				\clip (-50.000000, -36.000000) +(-45.000000:8.485281pt and 8.485281pt) -- +(45.000000:8.485281pt and 8.485281pt) -- +(135.000000:8.485281pt and 8.485281pt) -- +(225.000000:8.485281pt and 8.485281pt) -- cycle;
				\draw (-50.000000, -36.000000) node {$H$};
			\end{scope}
			% Line 12: b H
			\begin{scope}
				\draw[fill=white] (-25.000000, -36.000000) +(-45.000000:8.485281pt and 8.485281pt) -- +(45.000000:8.485281pt and 8.485281pt) -- +(135.000000:8.485281pt and 8.485281pt) -- +(225.000000:8.485281pt and 8.485281pt) -- cycle;
				\clip (-25.000000, -36.000000) +(-45.000000:8.485281pt and 8.485281pt) -- +(45.000000:8.485281pt and 8.485281pt) -- +(135.000000:8.485281pt and 8.485281pt) -- +(225.000000:8.485281pt and 8.485281pt) -- cycle;
				\draw (-25.000000, -36.000000) node {$H$};
			\end{scope}
			% Line 13: c H
			\begin{scope}
				\draw[fill=white] (0.000000, -36.000000) +(-45.000000:8.485281pt and 8.485281pt) -- +(45.000000:8.485281pt and 8.485281pt) -- +(135.000000:8.485281pt and 8.485281pt) -- +(225.000000:8.485281pt and 8.485281pt) -- cycle;
				\clip (0.000000, -36.000000) +(-45.000000:8.485281pt and 8.485281pt) -- +(45.000000:8.485281pt and 8.485281pt) -- +(135.000000:8.485281pt and 8.485281pt) -- +(225.000000:8.485281pt and 8.485281pt) -- cycle;
				\draw (0.000000, -36.000000) node {$H$};
			\end{scope}
			% Line 15: a G {}
			\begin{scope}
				\draw[fill=white] (-50.000000, -60.000000) +(-45.000000:8.485281pt and 8.485281pt) -- +(45.000000:8.485281pt and 8.485281pt) -- +(135.000000:8.485281pt and 8.485281pt) -- +(225.000000:8.485281pt and 8.485281pt) -- cycle;
				\clip (-50.000000, -60.000000) +(-45.000000:8.485281pt and 8.485281pt) -- +(45.000000:8.485281pt and 8.485281pt) -- +(135.000000:8.485281pt and 8.485281pt) -- +(225.000000:8.485281pt and 8.485281pt) -- cycle;
				\draw (-50.000000, -60.000000) node {{}};
			\end{scope}
			% Line 16: b G {}
			\begin{scope}
				\draw[fill=white] (-25.000000, -60.000000) +(-45.000000:8.485281pt and 8.485281pt) -- +(45.000000:8.485281pt and 8.485281pt) -- +(135.000000:8.485281pt and 8.485281pt) -- +(225.000000:8.485281pt and 8.485281pt) -- cycle;
				\clip (-25.000000, -60.000000) +(-45.000000:8.485281pt and 8.485281pt) -- +(45.000000:8.485281pt and 8.485281pt) -- +(135.000000:8.485281pt and 8.485281pt) -- +(225.000000:8.485281pt and 8.485281pt) -- cycle;
				\draw (-25.000000, -60.000000) node {{}};
			\end{scope}
			% Line 17: c G {}
			\begin{scope}
				\draw[fill=white] (0.000000, -60.000000) +(-45.000000:8.485281pt and 8.485281pt) -- +(45.000000:8.485281pt and 8.485281pt) -- +(135.000000:8.485281pt and 8.485281pt) -- +(225.000000:8.485281pt and 8.485281pt) -- cycle;
				\clip (0.000000, -60.000000) +(-45.000000:8.485281pt and 8.485281pt) -- +(45.000000:8.485281pt and 8.485281pt) -- +(135.000000:8.485281pt and 8.485281pt) -- +(225.000000:8.485281pt and 8.485281pt) -- cycle;
				\draw (0.000000, -60.000000) node {{}};
			\end{scope}
			% Line 19: a G {}
			\begin{scope}
				\draw[fill=white] (-50.000000, -84.000000) +(-45.000000:8.485281pt and 8.485281pt) -- +(45.000000:8.485281pt and 8.485281pt) -- +(135.000000:8.485281pt and 8.485281pt) -- +(225.000000:8.485281pt and 8.485281pt) -- cycle;
				\clip (-50.000000, -84.000000) +(-45.000000:8.485281pt and 8.485281pt) -- +(45.000000:8.485281pt and 8.485281pt) -- +(135.000000:8.485281pt and 8.485281pt) -- +(225.000000:8.485281pt and 8.485281pt) -- cycle;
				\draw (-50.000000, -84.000000) node {{}};
			\end{scope}
			% Line 20: b G {}
			\begin{scope}
				\draw[fill=white] (-25.000000, -84.000000) +(-45.000000:8.485281pt and 8.485281pt) -- +(45.000000:8.485281pt and 8.485281pt) -- +(135.000000:8.485281pt and 8.485281pt) -- +(225.000000:8.485281pt and 8.485281pt) -- cycle;
				\clip (-25.000000, -84.000000) +(-45.000000:8.485281pt and 8.485281pt) -- +(45.000000:8.485281pt and 8.485281pt) -- +(135.000000:8.485281pt and 8.485281pt) -- +(225.000000:8.485281pt and 8.485281pt) -- cycle;
				\draw (-25.000000, -84.000000) node {{}};
			\end{scope}
			% Line 21: c G {}
			\begin{scope}
				\draw[fill=white] (0.000000, -84.000000) +(-45.000000:8.485281pt and 8.485281pt) -- +(45.000000:8.485281pt and 8.485281pt) -- +(135.000000:8.485281pt and 8.485281pt) -- +(225.000000:8.485281pt and 8.485281pt) -- cycle;
				\clip (0.000000, -84.000000) +(-45.000000:8.485281pt and 8.485281pt) -- +(45.000000:8.485281pt and 8.485281pt) -- +(135.000000:8.485281pt and 8.485281pt) -- +(225.000000:8.485281pt and 8.485281pt) -- cycle;
				\draw (0.000000, -84.000000) node {{}};
			\end{scope}
			% Line 23: a C b
			\draw (-50.000000,-108.000000) -- (-25.000000,-108.000000);
			\begin{scope}
				\draw[fill=white] (-50.000000, -108.000000) circle(3.000000pt);
				\clip (-50.000000, -108.000000) circle(3.000000pt);
				\draw (-53.000000, -108.000000) -- (-47.000000, -108.000000);
				\draw (-50.000000, -111.000000) -- (-50.000000, -105.000000);
			\end{scope}
			\filldraw (-25.000000, -108.000000) circle(1.500000pt);
			% Line 24: c X
			\begin{scope}
				\draw[fill=white] (0.000000, -108.000000) +(-45.000000:8.485281pt and 8.485281pt) -- +(45.000000:8.485281pt and 8.485281pt) -- +(135.000000:8.485281pt and 8.485281pt) -- +(225.000000:8.485281pt and 8.485281pt) -- cycle;
				\clip (0.000000, -108.000000) +(-45.000000:8.485281pt and 8.485281pt) -- +(45.000000:8.485281pt and 8.485281pt) -- +(135.000000:8.485281pt and 8.485281pt) -- +(225.000000:8.485281pt and 8.485281pt) -- cycle;
				\draw (0.000000, -108.000000) node {$X$};
			\end{scope}
			% Line 26: a G {}
			\begin{scope}
				\draw[fill=white] (-50.000000, -132.000000) +(-45.000000:8.485281pt and 8.485281pt) -- +(45.000000:8.485281pt and 8.485281pt) -- +(135.000000:8.485281pt and 8.485281pt) -- +(225.000000:8.485281pt and 8.485281pt) -- cycle;
				\clip (-50.000000, -132.000000) +(-45.000000:8.485281pt and 8.485281pt) -- +(45.000000:8.485281pt and 8.485281pt) -- +(135.000000:8.485281pt and 8.485281pt) -- +(225.000000:8.485281pt and 8.485281pt) -- cycle;
				\draw (-50.000000, -132.000000) node {{}};
			\end{scope}
			% Line 27: b G {}
			\begin{scope}
				\draw[fill=white] (-25.000000, -132.000000) +(-45.000000:8.485281pt and 8.485281pt) -- +(45.000000:8.485281pt and 8.485281pt) -- +(135.000000:8.485281pt and 8.485281pt) -- +(225.000000:8.485281pt and 8.485281pt) -- cycle;
				\clip (-25.000000, -132.000000) +(-45.000000:8.485281pt and 8.485281pt) -- +(45.000000:8.485281pt and 8.485281pt) -- +(135.000000:8.485281pt and 8.485281pt) -- +(225.000000:8.485281pt and 8.485281pt) -- cycle;
				\draw (-25.000000, -132.000000) node {{}};
			\end{scope}
			% Line 28: c G {}
			\begin{scope}
				\draw[fill=white] (0.000000, -132.000000) +(-45.000000:8.485281pt and 8.485281pt) -- +(45.000000:8.485281pt and 8.485281pt) -- +(135.000000:8.485281pt and 8.485281pt) -- +(225.000000:8.485281pt and 8.485281pt) -- cycle;
				\clip (0.000000, -132.000000) +(-45.000000:8.485281pt and 8.485281pt) -- +(45.000000:8.485281pt and 8.485281pt) -- +(135.000000:8.485281pt and 8.485281pt) -- +(225.000000:8.485281pt and 8.485281pt) -- cycle;
				\draw (0.000000, -132.000000) node {{}};
			\end{scope}
			% Line 30: a G {}
			\begin{scope}
				\draw[fill=white] (-50.000000, -156.000000) +(-45.000000:8.485281pt and 8.485281pt) -- +(45.000000:8.485281pt and 8.485281pt) -- +(135.000000:8.485281pt and 8.485281pt) -- +(225.000000:8.485281pt and 8.485281pt) -- cycle;
				\clip (-50.000000, -156.000000) +(-45.000000:8.485281pt and 8.485281pt) -- +(45.000000:8.485281pt and 8.485281pt) -- +(135.000000:8.485281pt and 8.485281pt) -- +(225.000000:8.485281pt and 8.485281pt) -- cycle;
				\draw (-50.000000, -156.000000) node {{}};
			\end{scope}
			% Line 31: b G {}
			\begin{scope}
				\draw[fill=white] (-25.000000, -156.000000) +(-45.000000:8.485281pt and 8.485281pt) -- +(45.000000:8.485281pt and 8.485281pt) -- +(135.000000:8.485281pt and 8.485281pt) -- +(225.000000:8.485281pt and 8.485281pt) -- cycle;
				\clip (-25.000000, -156.000000) +(-45.000000:8.485281pt and 8.485281pt) -- +(45.000000:8.485281pt and 8.485281pt) -- +(135.000000:8.485281pt and 8.485281pt) -- +(225.000000:8.485281pt and 8.485281pt) -- cycle;
				\draw (-25.000000, -156.000000) node {{}};
			\end{scope}
			% Line 32: c G {}
			\begin{scope}
				\draw[fill=white] (0.000000, -156.000000) +(-45.000000:8.485281pt and 8.485281pt) -- +(45.000000:8.485281pt and 8.485281pt) -- +(135.000000:8.485281pt and 8.485281pt) -- +(225.000000:8.485281pt and 8.485281pt) -- cycle;
				\clip (0.000000, -156.000000) +(-45.000000:8.485281pt and 8.485281pt) -- +(45.000000:8.485281pt and 8.485281pt) -- +(135.000000:8.485281pt and 8.485281pt) -- +(225.000000:8.485281pt and 8.485281pt) -- cycle;
				\draw (0.000000, -156.000000) node {{}};
			\end{scope}
			% Line 34: a G {T}
			\begin{scope}
				\draw[fill=white] (-50.000000, -180.000000) +(-45.000000:8.485281pt and 8.485281pt) -- +(45.000000:8.485281pt and 8.485281pt) -- +(135.000000:8.485281pt and 8.485281pt) -- +(225.000000:8.485281pt and 8.485281pt) -- cycle;
				\clip (-50.000000, -180.000000) +(-45.000000:8.485281pt and 8.485281pt) -- +(45.000000:8.485281pt and 8.485281pt) -- +(135.000000:8.485281pt and 8.485281pt) -- +(225.000000:8.485281pt and 8.485281pt) -- cycle;
				\draw (-50.000000, -180.000000) node {{T}};
			\end{scope}
			% Line 35: b Z c
			\draw (-25.000000,-180.000000) -- (-0.000000,-180.000000);
			\begin{scope}
				\draw[fill=white] (-25.000000, -180.000000) +(-45.000000:8.485281pt and 8.485281pt) -- +(45.000000:8.485281pt and 8.485281pt) -- +(135.000000:8.485281pt and 8.485281pt) -- +(225.000000:8.485281pt and 8.485281pt) -- cycle;
				\clip (-25.000000, -180.000000) +(-45.000000:8.485281pt and 8.485281pt) -- +(45.000000:8.485281pt and 8.485281pt) -- +(135.000000:8.485281pt and 8.485281pt) -- +(225.000000:8.485281pt and 8.485281pt) -- cycle;
				\draw (-25.000000, -180.000000) node {$Z$};
			\end{scope}
			\filldraw (-0.000000, -180.000000) circle(1.500000pt);
			% Line 37: a G {}
			\begin{scope}
				\draw[fill=white] (-50.000000, -204.000000) +(-45.000000:8.485281pt and 8.485281pt) -- +(45.000000:8.485281pt and 8.485281pt) -- +(135.000000:8.485281pt and 8.485281pt) -- +(225.000000:8.485281pt and 8.485281pt) -- cycle;
				\clip (-50.000000, -204.000000) +(-45.000000:8.485281pt and 8.485281pt) -- +(45.000000:8.485281pt and 8.485281pt) -- +(135.000000:8.485281pt and 8.485281pt) -- +(225.000000:8.485281pt and 8.485281pt) -- cycle;
				\draw (-50.000000, -204.000000) node {{}};
			\end{scope}
			% Line 38: b G {}
			\begin{scope}
				\draw[fill=white] (-25.000000, -204.000000) +(-45.000000:8.485281pt and 8.485281pt) -- +(45.000000:8.485281pt and 8.485281pt) -- +(135.000000:8.485281pt and 8.485281pt) -- +(225.000000:8.485281pt and 8.485281pt) -- cycle;
				\clip (-25.000000, -204.000000) +(-45.000000:8.485281pt and 8.485281pt) -- +(45.000000:8.485281pt and 8.485281pt) -- +(135.000000:8.485281pt and 8.485281pt) -- +(225.000000:8.485281pt and 8.485281pt) -- cycle;
				\draw (-25.000000, -204.000000) node {{}};
			\end{scope}
			% Line 39: c G {}
			\begin{scope}
				\draw[fill=white] (0.000000, -204.000000) +(-45.000000:8.485281pt and 8.485281pt) -- +(45.000000:8.485281pt and 8.485281pt) -- +(135.000000:8.485281pt and 8.485281pt) -- +(225.000000:8.485281pt and 8.485281pt) -- cycle;
				\clip (0.000000, -204.000000) +(-45.000000:8.485281pt and 8.485281pt) -- +(45.000000:8.485281pt and 8.485281pt) -- +(135.000000:8.485281pt and 8.485281pt) -- +(225.000000:8.485281pt and 8.485281pt) -- cycle;
				\draw (0.000000, -204.000000) node {{}};
			\end{scope}
			% Line 41: a G {}
			\begin{scope}
				\draw[fill=white] (-50.000000, -228.000000) +(-45.000000:8.485281pt and 8.485281pt) -- +(45.000000:8.485281pt and 8.485281pt) -- +(135.000000:8.485281pt and 8.485281pt) -- +(225.000000:8.485281pt and 8.485281pt) -- cycle;
				\clip (-50.000000, -228.000000) +(-45.000000:8.485281pt and 8.485281pt) -- +(45.000000:8.485281pt and 8.485281pt) -- +(135.000000:8.485281pt and 8.485281pt) -- +(225.000000:8.485281pt and 8.485281pt) -- cycle;
				\draw (-50.000000, -228.000000) node {{}};
			\end{scope}
			% Line 42: b G {}
			\begin{scope}
				\draw[fill=white] (-25.000000, -228.000000) +(-45.000000:8.485281pt and 8.485281pt) -- +(45.000000:8.485281pt and 8.485281pt) -- +(135.000000:8.485281pt and 8.485281pt) -- +(225.000000:8.485281pt and 8.485281pt) -- cycle;
				\clip (-25.000000, -228.000000) +(-45.000000:8.485281pt and 8.485281pt) -- +(45.000000:8.485281pt and 8.485281pt) -- +(135.000000:8.485281pt and 8.485281pt) -- +(225.000000:8.485281pt and 8.485281pt) -- cycle;
				\draw (-25.000000, -228.000000) node {{}};
			\end{scope}
			% Line 43: c G {}
			\begin{scope}
				\draw[fill=white] (0.000000, -228.000000) +(-45.000000:8.485281pt and 8.485281pt) -- +(45.000000:8.485281pt and 8.485281pt) -- +(135.000000:8.485281pt and 8.485281pt) -- +(225.000000:8.485281pt and 8.485281pt) -- cycle;
				\clip (0.000000, -228.000000) +(-45.000000:8.485281pt and 8.485281pt) -- +(45.000000:8.485281pt and 8.485281pt) -- +(135.000000:8.485281pt and 8.485281pt) -- +(225.000000:8.485281pt and 8.485281pt) -- cycle;
				\draw (0.000000, -228.000000) node {{}};
			\end{scope}
			% Line 45: a H
			\begin{scope}
				\draw[fill=white] (-50.000000, -252.000000) +(-45.000000:8.485281pt and 8.485281pt) -- +(45.000000:8.485281pt and 8.485281pt) -- +(135.000000:8.485281pt and 8.485281pt) -- +(225.000000:8.485281pt and 8.485281pt) -- cycle;
				\clip (-50.000000, -252.000000) +(-45.000000:8.485281pt and 8.485281pt) -- +(45.000000:8.485281pt and 8.485281pt) -- +(135.000000:8.485281pt and 8.485281pt) -- +(225.000000:8.485281pt and 8.485281pt) -- cycle;
				\draw (-50.000000, -252.000000) node {$H$};
			\end{scope}
			% Line 46: b H
			\begin{scope}
				\draw[fill=white] (-25.000000, -252.000000) +(-45.000000:8.485281pt and 8.485281pt) -- +(45.000000:8.485281pt and 8.485281pt) -- +(135.000000:8.485281pt and 8.485281pt) -- +(225.000000:8.485281pt and 8.485281pt) -- cycle;
				\clip (-25.000000, -252.000000) +(-45.000000:8.485281pt and 8.485281pt) -- +(45.000000:8.485281pt and 8.485281pt) -- +(135.000000:8.485281pt and 8.485281pt) -- +(225.000000:8.485281pt and 8.485281pt) -- cycle;
				\draw (-25.000000, -252.000000) node {$H$};
			\end{scope}
			% Line 47: c H
			\begin{scope}
				\draw[fill=white] (0.000000, -252.000000) +(-45.000000:8.485281pt and 8.485281pt) -- +(45.000000:8.485281pt and 8.485281pt) -- +(135.000000:8.485281pt and 8.485281pt) -- +(225.000000:8.485281pt and 8.485281pt) -- cycle;
				\clip (0.000000, -252.000000) +(-45.000000:8.485281pt and 8.485281pt) -- +(45.000000:8.485281pt and 8.485281pt) -- +(135.000000:8.485281pt and 8.485281pt) -- +(225.000000:8.485281pt and 8.485281pt) -- cycle;
				\draw (0.000000, -252.000000) node {$H$};
			\end{scope}
			% Line 49: a G {}
			\begin{scope}
				\draw[fill=white] (-50.000000, -276.000000) +(-45.000000:8.485281pt and 8.485281pt) -- +(45.000000:8.485281pt and 8.485281pt) -- +(135.000000:8.485281pt and 8.485281pt) -- +(225.000000:8.485281pt and 8.485281pt) -- cycle;
				\clip (-50.000000, -276.000000) +(-45.000000:8.485281pt and 8.485281pt) -- +(45.000000:8.485281pt and 8.485281pt) -- +(135.000000:8.485281pt and 8.485281pt) -- +(225.000000:8.485281pt and 8.485281pt) -- cycle;
				\draw (-50.000000, -276.000000) node {{}};
			\end{scope}
			% Line 50: b G {}
			\begin{scope}
				\draw[fill=white] (-25.000000, -276.000000) +(-45.000000:8.485281pt and 8.485281pt) -- +(45.000000:8.485281pt and 8.485281pt) -- +(135.000000:8.485281pt and 8.485281pt) -- +(225.000000:8.485281pt and 8.485281pt) -- cycle;
				\clip (-25.000000, -276.000000) +(-45.000000:8.485281pt and 8.485281pt) -- +(45.000000:8.485281pt and 8.485281pt) -- +(135.000000:8.485281pt and 8.485281pt) -- +(225.000000:8.485281pt and 8.485281pt) -- cycle;
				\draw (-25.000000, -276.000000) node {{}};
			\end{scope}
			% Line 51: c G {}
			\begin{scope}
				\draw[fill=white] (0.000000, -276.000000) +(-45.000000:8.485281pt and 8.485281pt) -- +(45.000000:8.485281pt and 8.485281pt) -- +(135.000000:8.485281pt and 8.485281pt) -- +(225.000000:8.485281pt and 8.485281pt) -- cycle;
				\clip (0.000000, -276.000000) +(-45.000000:8.485281pt and 8.485281pt) -- +(45.000000:8.485281pt and 8.485281pt) -- +(135.000000:8.485281pt and 8.485281pt) -- +(225.000000:8.485281pt and 8.485281pt) -- cycle;
				\draw (0.000000, -276.000000) node {{}};
			\end{scope}
			% Line 53: a M
			\draw[fill=white] (-56.000000, -306.000000) rectangle (-44.000000, -294.000000);
			\draw[very thin] (-50.000000, -299.400000) arc (90:150:6.000000pt);
			\draw[very thin] (-50.000000, -299.400000) arc (90:30:6.000000pt);
			\draw[->,>=stealth] (-50.000000, -305.400000) -- +(80:10.392305pt);
			% Line 54: b M
			\draw[fill=white] (-31.000000, -306.000000) rectangle (-19.000000, -294.000000);
			\draw[very thin] (-25.000000, -299.400000) arc (90:150:6.000000pt);
			\draw[very thin] (-25.000000, -299.400000) arc (90:30:6.000000pt);
			\draw[->,>=stealth] (-25.000000, -305.400000) -- +(80:10.392305pt);
			% Line 55: c M
			\draw[fill=white] (-6.000000, -306.000000) rectangle (6.000000, -294.000000);
			\draw[very thin] (-0.000000, -299.400000) arc (90:150:6.000000pt);
			\draw[very thin] (-0.000000, -299.400000) arc (90:30:6.000000pt);
			\draw[->,>=stealth] (-0.000000, -305.400000) -- +(80:10.392305pt);
			% Done with gates; drawing ending labels
			% Done with ending labels; drawing cut lines and comments
			% Line 57: a b c @ 1 1 color=black style=dashed,rounded_corners=5pt
			\draw[draw opacity=1.000000,fill opacity=0.200000,color=black,dashed,rounded corners=5pt] (-62.500000,-27.000000) rectangle (12.500000,-45.000000);
			\draw[draw opacity=1.000000,fill opacity=0.200000,color=black,dashed,rounded corners=5pt] (-62.500000,-27.000000) rectangle (12.500000,-45.000000);
			% Line 58: a b c @ 4 4 color=black style=dashed,rounded_corners=5pt
			\draw[draw opacity=1.000000,fill opacity=0.200000,color=black,dashed,rounded corners=5pt] (-62.500000,-99.000000) rectangle (12.500000,-117.000000);
			\draw[draw opacity=1.000000,fill opacity=0.200000,color=black,dashed,rounded corners=5pt] (-62.500000,-99.000000) rectangle (12.500000,-117.000000);
			% Line 59: a b c @ 7 7 color=black style=dashed,rounded_corners=5pt
			\draw[draw opacity=1.000000,fill opacity=0.200000,color=black,dashed,rounded corners=5pt] (-62.500000,-171.000000) rectangle (12.500000,-189.000000);
			\draw[draw opacity=1.000000,fill opacity=0.200000,color=black,dashed,rounded corners=5pt] (-62.500000,-171.000000) rectangle (12.500000,-189.000000);
			% Line 60: a b c @ 10 10 color=black style=dashed,rounded_corners=5pt
			\draw[draw opacity=1.000000,fill opacity=0.200000,color=black,dashed,rounded corners=5pt] (-62.500000,-243.000000) rectangle (12.500000,-261.000000);
			\draw[draw opacity=1.000000,fill opacity=0.200000,color=black,dashed,rounded corners=5pt] (-62.500000,-243.000000) rectangle (12.500000,-261.000000);
			% Done with comments
		\end{tikzpicture}
		\caption{Circuit with randomized frames around each cycle. Arbitrary random gates are inserted before and after each cycle to ``frame'' the cycle.}
		\label{fig:framed-cycles}
	\end{figure}
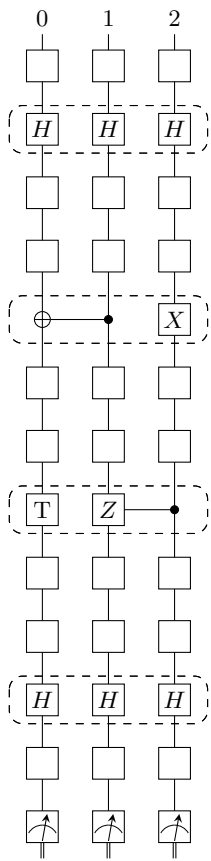

	Once the cycles have been defined, each cycle must be framed by inserting random single-qubit gates on either side of each qubit in the cycle. This is depicted in Figure \ref{fig:framed-cycles}. These random gates are chosen in such a way that the overall circuit unitary stays the same. For example, if a Hadamard is randomly chosen to frame one side of gate, another Hadamard is inserted on the other side of the gate. When a qubit is measured, we do not insert randomized gates for that cycle.
	
	At this step, the circuit has become much deeper, however because of the choice of randomized gates, the circuit can always be compiled to preserve overall circuit depth after compiling \cite{tketPaper}.
	
	Both universal frame randomization and Randomized Compilation can mitigate errors for circuits with a universal gate set. \cite{Wallman_2016}. Universal frame randomization allows cycles to contain any combination of single and two-qubit gates, while Randomized Compilation distinguishes between these types of gates and does not allow two qubit gates inside cycles.
	
	We will investigate universal frame randomization by using the TKET software package and Randomized Compilation by using the True-Q software package.
	
	\newpage
	
	\subsection{The Ising Model}
	
	The Ising model describes a system of connected nodes that can be in one of two states. We can describe the energy of this system with the following Hamiltonian where $J$ is a matrix representing interactions between nodes and $h$ represents an external magnetic field
	
	\begin{equation}\label{ham}
		\mathcal{H} = - \sum_{\braket{i,j}} J_{i,j} \sigma_i \sigma_j -  \sum_i h \sigma_i.
	\end{equation}
	
	In this model, frustration occurs when at least one edge must be violated, meaning the state chosen for the two adjacent nodes do not minimize the energy of that coupling term.
	
	\begin{figure}[H]
		\centering
		\includegraphics[width=0.45\textwidth]{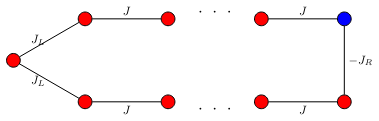}
		\caption{Frustrated Ising ring \cite{Roberts_2020}. Nodes are connected by coupling terms denoted $J$, $J_L$, and $-J_R$. The negative coupling term causes frustration. Red nodes are in one state (spin up) and the blue node is in the other state (spin down).}
		\label{fig:isingRing}
	\end{figure}
	
	This can occur in a ring where all but one of the couplings are positive, but the last coupling is negative as shown in Figure \ref{fig:isingRing}. In the lowest energy state of this system, the nodes must choose states where they do not agree with the sign of at least one of the couplings. Often frustration can lead to a system with multiple distinct ground states. 
	
	The frustrated Ising model poses significant challenges for quantum annealers, particularly for certain parameter choices \cite{Roberts_2020}. This motivated our investigation into the performance of QAOA on this problem.
	
	\subsection{Quantum Approximate Optimization Algorithm (QAOA)}
	
	The Quantum Approximate Optimization Algorithm, or QAOA, is a hybrid quantum-classical algorithm that can be used to find approximate solutions to combinatorial optimization problems \cite{farhi2014quantum}\cite{Zhou_2020}. It is a discrete approximation of quantum annealing that can be run on gate-based quantum computers \cite{farhi2000quantum}.
	
\begin{figure}[]
		\centering
		\begin{tikzpicture}[scale=1.000000,x=1pt,y=1pt]
			\filldraw[color=white] (12.500000, 0.000000) rectangle (-87.500000, -303.000000);
			% Drawing wires
			% Line 2: a W 0 width=25
			\draw[color=black] (-75.000000,0.000000) -- (-75.000000,-291.000000);
			\draw[color=black] (-74.500000,-291.000000) -- (-74.500000,-303.000000);
			\draw[color=black] (-75.500000,-291.000000) -- (-75.500000,-303.000000);
			\draw[color=black] (-75.000000,0.000000) node[above] {$0$};
			% Line 3: b W 1 width=25
			\draw[color=black] (-50.000000,0.000000) -- (-50.000000,-291.000000);
			\draw[color=black] (-49.500000,-291.000000) -- (-49.500000,-303.000000);
			\draw[color=black] (-50.500000,-291.000000) -- (-50.500000,-303.000000);
			\draw[color=black] (-50.000000,0.000000) node[above] {$1$};
			% Line 4: c W 2 width=25
			\draw[color=black] (-25.000000,0.000000) -- (-25.000000,-291.000000);
			\draw[color=black] (-24.500000,-291.000000) -- (-24.500000,-303.000000);
			\draw[color=black] (-25.500000,-291.000000) -- (-25.500000,-303.000000);
			\draw[color=black] (-25.000000,0.000000) node[above] {$2$};
			% Line 5: d W 3 width=25
			\draw[color=black] (-0.000000,0.000000) -- (-0.000000,-291.000000);
			\draw[color=black] (0.500000,-291.000000) -- (0.500000,-303.000000);
			\draw[color=black] (-0.500000,-291.000000) -- (-0.500000,-303.000000);
			\draw[color=black] (-0.000000,0.000000) node[above] {$3$};
			% Done with wires; drawing gates
			% Line 7: a b c d G {$e^{-i \gamma_1 \mathcal{H}_C}$} height=30
			\draw (-75.000000,-21.000000) -- (-0.000000,-21.000000);
			\begin{scope}
				\draw[fill=white] (-37.500000, -21.000000) +(-45.000000:61.518290pt and 21.213203pt) -- +(45.000000:61.518290pt and 21.213203pt) -- +(135.000000:61.518290pt and 21.213203pt) -- +(225.000000:61.518290pt and 21.213203pt) -- cycle;
				\clip (-37.500000, -21.000000) +(-45.000000:61.518290pt and 21.213203pt) -- +(45.000000:61.518290pt and 21.213203pt) -- +(135.000000:61.518290pt and 21.213203pt) -- +(225.000000:61.518290pt and 21.213203pt) -- cycle;
				\draw (-37.500000, -21.000000) node {{$e^{-i \gamma_1 \mathcal{H}_C}$}};
			\end{scope}
			% Line 8: a b c d G {$e^{-i \beta_1 \mathcal{H}_M}$} height=30
			\draw (-75.000000,-63.000000) -- (-0.000000,-63.000000);
			\begin{scope}
				\draw[fill=white] (-37.500000, -63.000000) +(-45.000000:61.518290pt and 21.213203pt) -- +(45.000000:61.518290pt and 21.213203pt) -- +(135.000000:61.518290pt and 21.213203pt) -- +(225.000000:61.518290pt and 21.213203pt) -- cycle;
				\clip (-37.500000, -63.000000) +(-45.000000:61.518290pt and 21.213203pt) -- +(45.000000:61.518290pt and 21.213203pt) -- +(135.000000:61.518290pt and 21.213203pt) -- +(225.000000:61.518290pt and 21.213203pt) -- cycle;
				\draw (-37.500000, -63.000000) node {{$e^{-i \beta_1 \mathcal{H}_M}$}};
			\end{scope}
			% Line 10: a b c d G {$e^{-i \gamma_2 \mathcal{H}_C}$} height=30
			\draw (-75.000000,-105.000000) -- (-0.000000,-105.000000);
			\begin{scope}
				\draw[fill=white] (-37.500000, -105.000000) +(-45.000000:61.518290pt and 21.213203pt) -- +(45.000000:61.518290pt and 21.213203pt) -- +(135.000000:61.518290pt and 21.213203pt) -- +(225.000000:61.518290pt and 21.213203pt) -- cycle;
				\clip (-37.500000, -105.000000) +(-45.000000:61.518290pt and 21.213203pt) -- +(45.000000:61.518290pt and 21.213203pt) -- +(135.000000:61.518290pt and 21.213203pt) -- +(225.000000:61.518290pt and 21.213203pt) -- cycle;
				\draw (-37.500000, -105.000000) node {{$e^{-i \gamma_2 \mathcal{H}_C}$}};
			\end{scope}
			% Line 11: a b c d G {$e^{-i \beta_2 \mathcal{H}_M}$} height=30
			\draw (-75.000000,-147.000000) -- (-0.000000,-147.000000);
			\begin{scope}
				\draw[fill=white] (-37.500000, -147.000000) +(-45.000000:61.518290pt and 21.213203pt) -- +(45.000000:61.518290pt and 21.213203pt) -- +(135.000000:61.518290pt and 21.213203pt) -- +(225.000000:61.518290pt and 21.213203pt) -- cycle;
				\clip (-37.500000, -147.000000) +(-45.000000:61.518290pt and 21.213203pt) -- +(45.000000:61.518290pt and 21.213203pt) -- +(135.000000:61.518290pt and 21.213203pt) -- +(225.000000:61.518290pt and 21.213203pt) -- cycle;
				\draw (-37.500000, -147.000000) node {{$e^{-i \beta_2 \mathcal{H}_M}$}};
			\end{scope}
			% Line 13: a b c d LABEL \cdots
			\draw[color=black] (-75.000000, -181.500000) node [fill=white, rotate around={-90:(0,0)}] {$\cdots$};
			\draw[color=black] (-50.000000, -181.500000) node [fill=white, rotate around={-90:(0,0)}] {$\cdots$};
			\draw[color=black] (-25.000000, -181.500000) node [fill=white, rotate around={-90:(0,0)}] {$\cdots$};
			\draw[color=black] (-0.000000, -181.500000) node [fill=white, rotate around={-90:(0,0)}] {$\cdots$};
			% Line 16: a b c d G {$e^{-i \gamma_p \mathcal{H}_C}$} height=30
			\draw (-75.000000,-216.000000) -- (-0.000000,-216.000000);
			\begin{scope}
				\draw[fill=white] (-37.500000, -216.000000) +(-45.000000:61.518290pt and 21.213203pt) -- +(45.000000:61.518290pt and 21.213203pt) -- +(135.000000:61.518290pt and 21.213203pt) -- +(225.000000:61.518290pt and 21.213203pt) -- cycle;
				\clip (-37.500000, -216.000000) +(-45.000000:61.518290pt and 21.213203pt) -- +(45.000000:61.518290pt and 21.213203pt) -- +(135.000000:61.518290pt and 21.213203pt) -- +(225.000000:61.518290pt and 21.213203pt) -- cycle;
				\draw (-37.500000, -216.000000) node {{$e^{-i \gamma_p \mathcal{H}_C}$}};
			\end{scope}
			% Line 17: a b c d G {$e^{-i \beta_p \mathcal{H}_M}$} height=30
			\draw (-75.000000,-258.000000) -- (-0.000000,-258.000000);
			\begin{scope}
				\draw[fill=white] (-37.500000, -258.000000) +(-45.000000:61.518290pt and 21.213203pt) -- +(45.000000:61.518290pt and 21.213203pt) -- +(135.000000:61.518290pt and 21.213203pt) -- +(225.000000:61.518290pt and 21.213203pt) -- cycle;
				\clip (-37.500000, -258.000000) +(-45.000000:61.518290pt and 21.213203pt) -- +(45.000000:61.518290pt and 21.213203pt) -- +(135.000000:61.518290pt and 21.213203pt) -- +(225.000000:61.518290pt and 21.213203pt) -- cycle;
				\draw (-37.500000, -258.000000) node {{$e^{-i \beta_p \mathcal{H}_M}$}};
			\end{scope}
			% Line 19: a b c d M
			\draw[fill=white] (-81.000000, -297.000000) rectangle (-69.000000, -285.000000);
			\draw[very thin] (-75.000000, -290.400000) arc (90:150:6.000000pt);
			\draw[very thin] (-75.000000, -290.400000) arc (90:30:6.000000pt);
			\draw[->,>=stealth] (-75.000000, -296.400000) -- +(80:10.392305pt);
			\draw[fill=white] (-56.000000, -297.000000) rectangle (-44.000000, -285.000000);
			\draw[very thin] (-50.000000, -290.400000) arc (90:150:6.000000pt);
			\draw[very thin] (-50.000000, -290.400000) arc (90:30:6.000000pt);
			\draw[->,>=stealth] (-50.000000, -296.400000) -- +(80:10.392305pt);
			\draw[fill=white] (-31.000000, -297.000000) rectangle (-19.000000, -285.000000);
			\draw[very thin] (-25.000000, -290.400000) arc (90:150:6.000000pt);
			\draw[very thin] (-25.000000, -290.400000) arc (90:30:6.000000pt);
			\draw[->,>=stealth] (-25.000000, -296.400000) -- +(80:10.392305pt);
			\draw[fill=white] (-6.000000, -297.000000) rectangle (6.000000, -285.000000);
			\draw[very thin] (-0.000000, -290.400000) arc (90:150:6.000000pt);
			\draw[very thin] (-0.000000, -290.400000) arc (90:30:6.000000pt);
			\draw[->,>=stealth] (-0.000000, -296.400000) -- +(80:10.392305pt);
			% Done with gates; drawing ending labels
			% Done with ending labels; drawing cut lines and comments
			% Line 21: a b c d @ 0 1 color=black style=dashed,rounded_corners=5pt
			\draw[draw opacity=1.000000,fill opacity=0.200000,color=black,dashed,rounded corners=5pt] (-87.500000,-3.000000) rectangle (12.500000,-81.000000);
			\draw[draw opacity=1.000000,fill opacity=0.200000,color=black,dashed,rounded corners=5pt] (-87.500000,-3.000000) rectangle (12.500000,-81.000000);
			% Done with comments
		\end{tikzpicture}
		\caption{Circuit diagram for QAOA. Gates parameterized by $\beta$ and $\gamma$ are iterated over the length of the circuit. The first complete layer of the circuit is circled with a dashed line.}
		\label{fig:qaoacircuit}
	\end{figure}
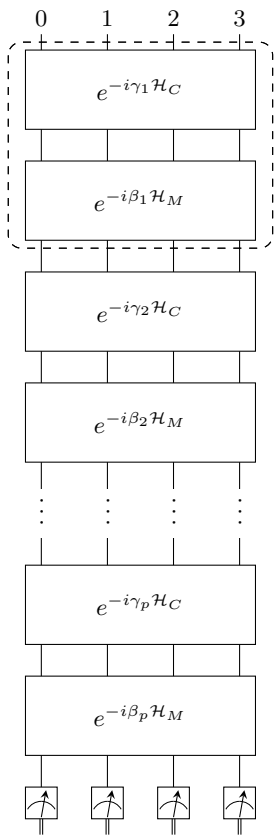

	A QAOA circuit consists of $p$ layers. In each layer, we apply a phase operator parameterized by $\gamma$ and then a mixing operator parameterized by $\beta$ \cite{Zhou_2020}. An example QAOA circuit is shown in Figure \ref{fig:qaoacircuit}. This circuit prepares a state as shown below where $\mathcal{H}_C$ represents the cost Hamiltonian and $\mathcal{H}_M$ represents the mixer Hamiltonian.
	
	$$\ket{\Psi(t)} = e^{-i\beta \mathcal{H}_C} e^{-i\gamma \mathcal{H}_M} \dots  e^{-i\beta \mathcal{H}_C} e^{-i\gamma \mathcal{H}_M} \ket{\Psi(0)}$$

	Here, we will set $p=1$ and evaluate the circuit for a range of $\gamma$ and $\beta$ hyper-parameters. This will allow us to examine errors. We can create a map of the expected energy values obtained from running each pair of $\gamma$ and $\beta$ values. We will refer to this as the energy landscape.
	
	\newpage
	
	\section{Experimental Setup}
	We investigate the use of QAOA to solve for the lowest energy state of a 12-node frustrated Ising ring. 
	To simplify the problem and focus on the coupling terms, we can set all $h$ terms to zero. Then the Hamiltonian in Equation \ref{ham} is determined only by the coupling terms. To model a ring, we choose adjacent coupling terms (i.e. between nodes (0,1), (1,2), (2,3), etc) to be 1, and all other coupling terms to be zero. Furthermore, to introduce frustration to the ring, we set one of these pairs to have a coupling term of -1. 
	
	We choose to focus on frustrated rings of 12 nodes because this matches the heavy hexagon layout of IBM devices \cite{ibmre}. When running the algorithm, each node in the problem is mapped directly to a qubit on the quantum computer, so the ring of 12 nodes perfectly matches the connectivity of many of IBM's devices \cite{qiskitdocs}. 
	
	When running QAOA to solve this problem, we choose to use only one layer, i.e. one phase operator and one mixing operator, and create an energy landscape over a range of gammas and betas. This choice allowed us to focus on analyzing errors.	
	
	In this case, the QAOA layer consists of Hadamard gates on each qubit, RZZ gates parameterized by $\gamma$ on each pair of qubits, then RX gates parameterized by $\beta$ on each qubit. Finally, the state of each qubit is measured. The total circuit is shown in the Appendix. 
	Notice that the 12 entangling RZZ gates are implemented in two cycles: first, RZZ gates are applied to qubits (0,1), (2,3), (4,5), (6,7), (8,9), (10,11); then RZZ gates are applied to qubits (0,11), (1,2), (3,4), (5,6), (7,8), (9,10). This minimizes circuit depth.
	
	To generate an energy landscape, we then iterate over a range of 17 $\beta$ values and 17 $\gamma$ values, each equally spaced on the interval [0,1]. This yields 289 unique $\gamma, \beta$ pairs. Applying these parameters, we generate 289 circuits and ran each on the quantum computer for 5000 shots, meaning that we ran each circuit 5000 times. 
	
	After all circuits are run, we use the results from each circuit to calculate the expected lowest energy value of the frustrated Ising ring.
	
	To compare errors, we run these circuits on a closed/ideal classical simulator, a noisy simulator that used a device-specific noise model, and on quantum processors. We then apply universal frame randomization and Randomized Compilation and run the modified circuits on quantum processors.
	
	\subsection{Randomized Compilation with True-Q}
	Using the True-Q package, we implement our circuit and define it to contain two randomized cycles because the entangling RZZ gates were implemented in two cycles.	
	We generate randomized circuits and run them on IBM quantum computers. 
	
	\subsection{Universal Frame Randomization with TKET}
	We also use TKET to implement universal frame randomization. In the same way, we build our circuit and define it to contain two cycles. We generate randomized circuits and run them on IBM quantum computers. A circuit generated using universal frame randomization is shown in Figure 10 in the Appendix. 
	
	\section{Results}
	For each collection of circuits, we generated an energy landscape plot to map the calculated energies for the range of $\gamma$ and $\beta$ parameters. 

\begin{figure}[H]
		\centering
		\includegraphics[width=0.4\textwidth]{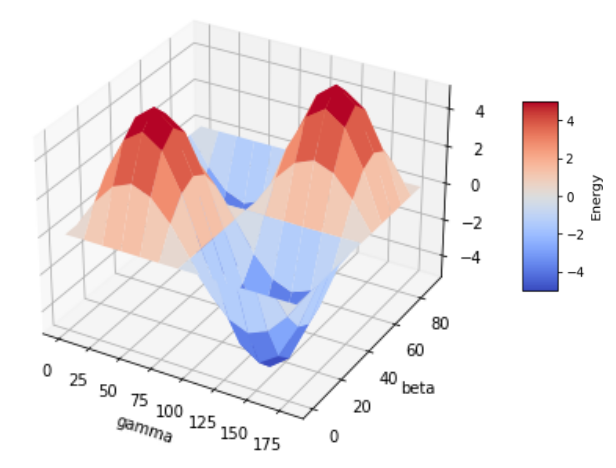}
		\caption{Energy Landscape generated by running QAOA on a closed IBM quantum simulator with no noise model (Fall 2022).}
		\label{fig:closedSim}
	\end{figure}

	First, we run QAOA with $p=1$ on a closed system. The energy landscape from this simulation is plotted in Figure \ref{fig:closedSim}.

\begin{figure}[H]
		\centering
		\includegraphics[width=0.41\textwidth]{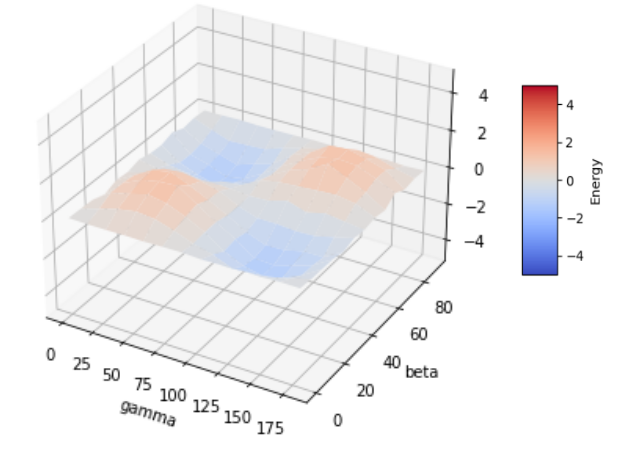}
		\caption{Energy Landscape generated by running QAOA on ibm\_mumbai without randomized compilation (Fall 2022).}
		\label{fig:noComps}
	\end{figure}

	We also run the same collection of circuits on a quantum processor with no additional optimizations or error mitigation (aside from the default compilation optimizations applied by IBM) This circuit produced noticeably lower absolute energy values as depicted in Figure \ref{fig:noComps}; however, it still demonstrates the same periodic behavior as the closed system. 
	
\begin{figure}[H]
		\centering
		\includegraphics[width=0.4\textwidth]{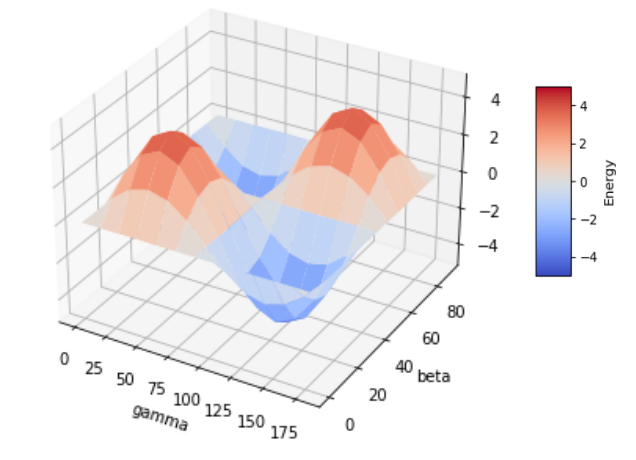}
		\caption{Energy Landscape generated by running QAOA on a noisy simulator of the ibm\_hanoi device without randomized compilation (Fall 2022).}
		\label{fig:noisy}
	\end{figure}

	Furthermore, we run the same QAOA circuits on a noisy simulator model. We choose a device-specific noise model, meaning that the classical simulator replicates the types and frequency of errors expected from a specific IBM device. As shown in Figure \ref{fig:noisy}, the energy landscape appears to show higher absolute energy values than the quantum processor but lower than the closed system, indicating that the noisy simulator effectively models some but not all of the errors of the device. 
	
	\begin{figure}[H]
		\centering
		\includegraphics[width=0.38\textwidth]{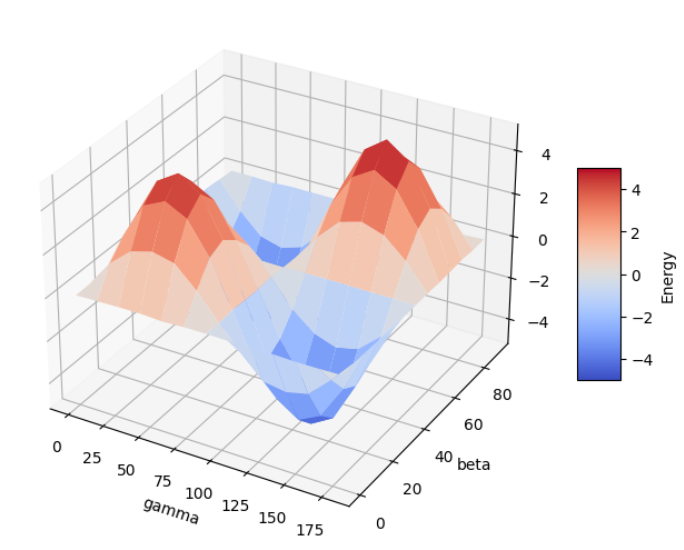}
		\caption{Energy Landscape generated by running QAOA on ibm\_hanoi using True-Q Randomized Compilation with 20 compilations (Fall 2022 - Spring 2023).}
		\label{fig:true-q20}
	\end{figure}

	\begin{figure}[H]
		\centering
		\includegraphics[width=.4 \textwidth]{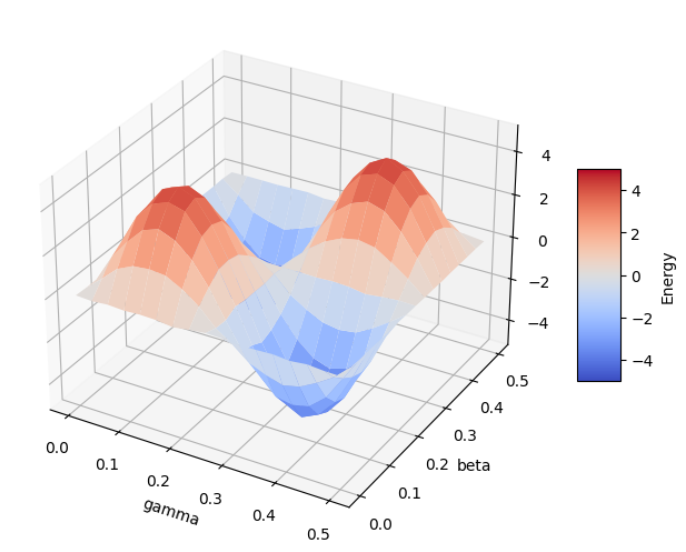}
		\caption{Energy Landscape generated by running QAOA on ibm\_hanoi with 10 compilations using TKET universal frame randomization (Spring 2024). }
		\label{fig:tket10}
	\end{figure}

	\subsection*{Randomized Compilation and Universal Frame Randomization}
	We apply Randomized Compilation and universal frame randomization to the same collection of circuits as before and compare the energy landscapes generated. We choose to apply 20 compilations because this was expected to provide a reasonable improvement in results without too much additional computation \cite{Wallman_2016}. 
	Results generated using True-Q are plotted in Figure \ref{fig:true-q20}. Results generated using TKET are plotted in Figure \ref{fig:tket10}.

	To compare these techniques, we extract the maximum and minimum energies from each plot and compare them to the results from a closed system simulator. We use Bayesian bootstrapping to estimate the standard deviation of these values, which are summarized in TABLE \ref{minMaxEng}. Note that this analysis focuses on the extremal values and may not capture all possible errors, as some may alter the energy landscape without affecting the minimum and maximum values.
	
	Also note that due to a limited trial license of the True-Q software, the True-Q and TKET experiments were run at different times. Approximate dates are included in the figure captions.
	
	\begin{table}[H]
		\vspace{8pt}
		\begin{tabular}{|c|c|c|}
			\hline
			Technique & Compilations & Extremal Energy \\
			\hline
			\hline
			Simulator & 1  & $5.676 \pm 0.006$ \\
			\hline
			Noisy Simulator & 1  & $2.92 \pm 0.107$ \\
			\hline
			No Error Mitigation & 1 & $2.63 \pm 0.068$ \\
			\hline
			Randomized & & \\Compilation & 20  &$ 5.25 \pm 0.145$ \\
			\hline
			Universal Frame & &\\Randomization & 20  & $4.08 \pm 0.36$ \\
			\hline
		\end{tabular}
		\caption{Absolute value of extremal energies with 2 standard deviations (Fall 2023 - Spring 2024). All results here are from IBM Falcon devices or IBM simulators. }
		\label{minMaxEng}
	\end{table}

	\section{Discussion}
	We use QAOA to solve for the lowest energy state of a 12-node frustrated Ising ring system, implementing it on a classical quantum simulator and IBM's superconducting quantum computers. On IBM's devices, we run these circuits with no error mitigation, using True-Q Randomized Compilation, and using TKET universal frame randomization. Across all runs, the quantum computer produces the same periodic pattern of high and low energy values in the energy landscape, although the simulator achieves a larger range of expected energy values due to its lack of noise. We observe a significant improvement in expected energy values when applying True-Q Randomized Compilation and TKET universal frame randomization, as shown in Figures \ref{fig:closedSim}-\ref{fig:tket10}.
	
	We conclude that universal frame randomization and Randomized Compilation show promise for minimizing errors in QAOA circuits, producing energy expectation values closer to those of the noiseless circuit than unmitigated circuits.
	
	In the future, we plan to explore the combined effects of additional optimization techniques with universal frame randomization or Randomized Compilation on the resulting energy values. We also intend to investigate how different numbers of random compilations impact the output.
	
	\vspace{12px}
	\begin{acknowledgments}
		This paper is adapted from an undergraduate thesis submitted to the Glynn Family Honors Program at the University of Notre Dame. We acknowledge the contributions of Anthony Hoffman, advisor for Rachel's thesis at Notre Dame, Jarrett Smalley, initial code developer, Arnaud Carignan-Dugas of Keysight Technologies for assistance with the True-Q software, and Joseph Emerson for feedback on the first version of this paper. We also thank Lockheed Martin for funding this project.
	\end{acknowledgments}
	
	\bibliography{thesis}
	
	\onecolumngrid
	\vspace{20cm}
	
	\section*{Appendix}
	\begin{figure}[H]
		\centering
		\includegraphics[width=0.95\textwidth]{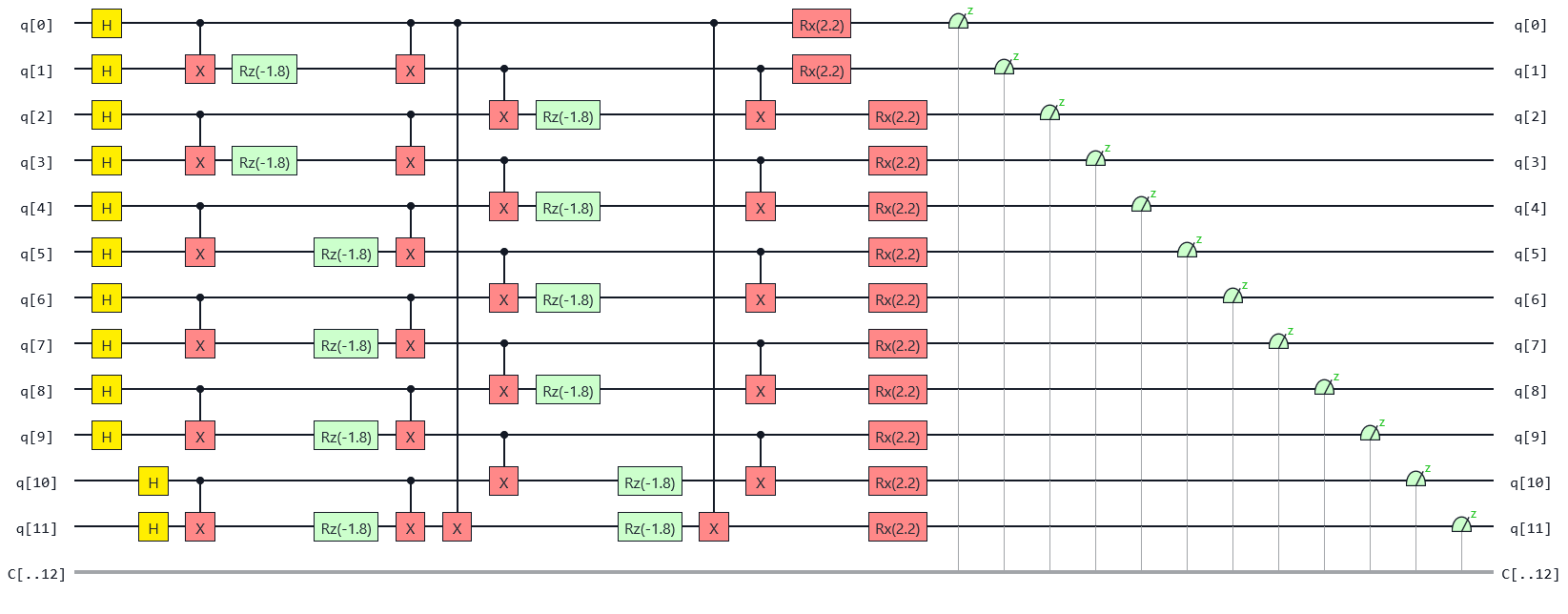}
		\caption{Circuit diagram generated using TKET with no randomized compilation. To implement QAOA to solve the frustrated Ising ring problem, we apply Hadamard gates, entangling RZZ gates (equivalent to CNOT, RZ, CNOT), and then RX gates before measuring the state of each qubit.}
		\label{fig:actualcircuit}
	\end{figure}
	
	\begin{figure}[H]
		\centering
		\includegraphics[width=\textwidth]{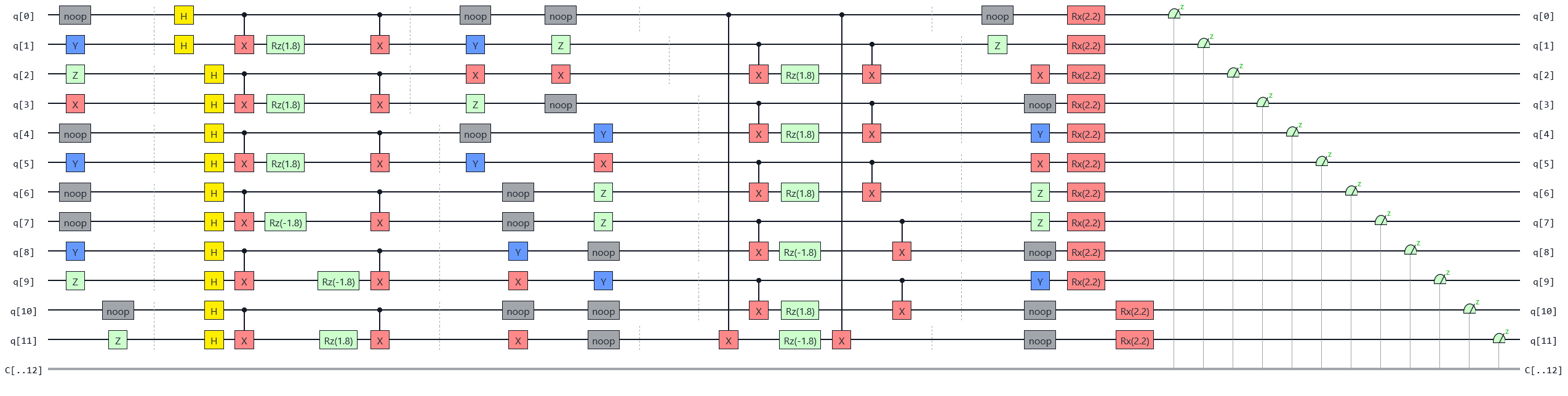}
		\caption{Randomized circuit diagram generated using TKET. The first cycle contains Hadamard gates and the first half of the entangling gates. The second cycle contains the second half of the entangling gates.}
		\label{fig:randomcircuit}
	\end{figure}

\end{document}